\documentclass[10pt,journal,compsoc]{IEEEtran}
\usepackage{mathtools,amssymb,lipsum}

\usepackage{cuted}
\usepackage{dsfont}
\usepackage{amsmath}
\usepackage{msc}
\usepackage{tikz}
\usetikzlibrary{shapes,arrows}
\usepackage{tcolorbox}
\usepackage{ amssymb }
\usepackage{balance}
\usepackage{graphicx}
\usepackage[normalem]{ulem}
\usepackage{url,cite}
\usepackage{balance}
\usepackage{graphicx}
\usepackage{enumerate}
\usepackage{stmaryrd}
\usepackage{color}
\usepackage{multirow}
\usepackage{array}
\usepackage{float}
\usepackage[english]{babel}
\usepackage[utf8]{inputenc}
\usepackage{algorithm}
\usepackage[noend]{algpseudocode}
\newcolumntype{L}[1]{>{\raggedright\let\newline\\\arraybackslash\hspace{0pt}}m{#1}}
\newcolumntype{C}[1]{>{\centering\let\newline\\\arraybackslash\hspace{0pt}}m{#1}}
\newcolumntype{R}[1]{>{\raggedleft\let\newline\\\arraybackslash\hspace{0pt}}m{#1}}

\usepackage{amsmath}

\IEEEtitleabstractindextext{%
\begin{abstract}
Recent advances in machine learning techniques  are enabling Automated Speech Recognition (ASR) more accurate and practical. The evidence of this can be seen in the rising number of smart devices with voice processing capabilities. More and more devices around us are in-built with ASR technology. This  poses serious privacy threats as speech contains unique biometric characteristics and personal data. However, the privacy concern can be mitigated if the voice features are processed in the encrypted domain. Within this context, this paper proposes an algorithm to redesign the back-end of the speaker verification system using fully homomorphic encryption techniques. The solution exploits the Cheon-Kim-Kim-Song (CKKS) fully homomorphic encryption scheme to obtain a real-time and non-interactive solution. The proposed solution contains a novel approach based on  Newton-Raphson method to overcome the limitation of CKKS scheme (i.e., calculating an inverse square-root of an encrypted number). This provides an efficient solution with less multiplicative depths for a negligible loss in accuracy. The proposed algorithm is validated using a well-known speech dataset. The proposed algorithm performs encrypted-domain verification in real-time (with less than 1.3 seconds delay) for a 2.8\% equal-error-rate loss compared to plain-domain verification.
\end{abstract}

}

\author{Yogachandran Rahulamathavan%, Kanapathippillai Cumanan, and Ahmet Kondoz
\IEEEcompsocitemizethanks{ 
\IEEEcompsocthanksitem The sourcecode of this paper can be found here https://github.com/rahulay1/iVectorTenSEAL/tree/master.
\IEEEcompsocthanksitem Y. Rahulamathavan is with the Institute for Digital Technologies, Loughborough University London, London, U.K. (e-mail: y.rahulamathavan@lboro.ac.uk).
\IEEEcompsocthanksitem \textcolor{red}{This work has been submitted to the IEEE for possible publication. Copyright may be transferred without notice, after which this version may no longer be accessible.}

}
}

\begin{document}

% \begin{frontmatter}
% \fntext[label3]{Mauro Conti is supported by a EU Marie Curie Fellowship for the project PRISM-CODE (grant n. PCIG11-GA-2012-321980). This work has been partially supported by the TENACE PRIN Project (grant n. 20103P34XC) funded by the Italian MIUR.}
%%%%%%%%%%%%%%%%%%%%%%%%%%%%%%%%%%%%%%%%%%%%%%%%%%%%%%%%%%%%%%%%%%%%%%%%%%%%%%%%

%\title{Privacy-preserving Speaker Verification Systems Using Lattice-based Cryptography}

\title{Privacy-preserving Similarity Calculation of Speaker Features Using Fully Homomorphic Encryption}
\maketitle

% \author[City]{Fei Li}
% \ead{fei.li.1@city.ac.uk}
% \author[City]{Yogachandran Rahulamathavan\corref{cor1}}
% \ead{yogachandran.rahulamathavan.1@city.ac.uk}
% \cortext[cor1]{Corresponding author. Tel.: +44 (0)20 7040 8377, Fax.: +44 (0)20 7040 8566}
% \author[Padova]{Mauro Conti}
% \ead{conti@math.unipd.it}
% \author[City]{Muttukrishnan Rajarajan}
% \ead{R.Muttukrishnan@city.ac.uk}
% %\author[Padova]{Mauro Conti}

% \address[City]{School of Engineering and Mathematical Science, City University London,  London, United Kingdom.}

% \address[Padova]{Department of Mathematics, University of Padova, Padova, Italy.}

\section{Introduction}\label{Section: Intro}
\begin{figure*}\centering
  \includegraphics[trim={0cm 0 0cm 0},clip,width=7in]{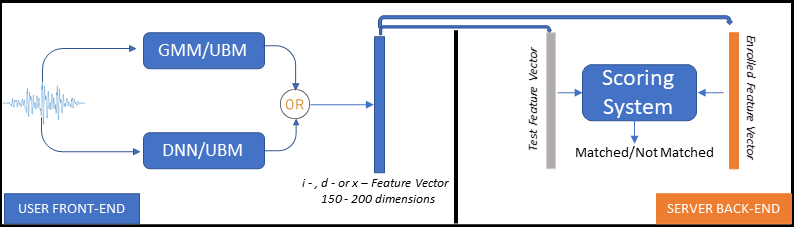}
\caption{Block diagram showing front-end and back-end of a typical speaker verification system. }
\label{Fig: BlockDiagram}
\end{figure*}

Automatic Speech Recognition (ASR) plays a major role in several emerging smart applications and services. Recent studies show that ASR can be used to detect emerging medical conditions such as Parkinson’s disease \cite{Parkinson},  Post-Traumatic Stress Disorder (PTSD) \cite{PTSD} and neurodegenerative diseases such as Alzheimer’s disease \cite{Alzheimer} and dementia \cite{Dementia} by continuously and passively observing the user's speech. ASR is also used in banking and financial sector for biometric verification purposes \cite{JainAK}. Moreover, several smart devices (i.e., smart TVs and speakers) are now embedded with ASR functionality. The commonality across all of these applications and services is that they all require user speech features to be sent to servers (or cloud) for classification purposes. These speech features are fed into machine learning models and matched to a known class. Fig. \ref{Fig: BlockDiagram} shows a typical ASR system. If the ASR is for speaker verification then the user's feature is matched against the  enrolled identity of the claimed speaker.

While these technologies are very useful for healthcare monitoring, to enhance security in banking and finance and to improve user experience, continuously sending the speech features to servers  pose serious privacy threats to the users. The users can be tracked by the service providers or their medical conditions can be inferred and sold to insurance companies or the speech biometrics can be hacked by adversaries. These are irrevocable problems. Within this context, this paper develops a privacy-preserving solution to redesign the back-end of the speaker verification system where user's speech features  remain in encrypted domain during the transmission, storage and processing.

Speaker verification is a task of verifying a user using their voice biometrics. As shown in Fig. \ref{Fig: BlockDiagram}, the speaker verification has two parts: 1) enrolment and 2) matching. During the first stage, the user needs to enrol their speech biometrics via speaker enrolment process. These enrolled biometric templates might be stored in an authentication server (which resides along with other servers). During the verification, a fresh speech feature is extracted and sent to the server who performs a comparison against the stored template using a machine learning technique. If the comparison is successful then the authentication server allows the user to access the service.

Traditionally the user's speech features (or the templates) are only encrypted during the storage but decrypted during the processing (i.e., verification) stage. This means that the server (or the adversaries who compromise the server) can access the features and track the users. This is where the privacy risk and this paper develops a novel technique to transforms the back-end processing in the encrypted domain. This paper proposes a technique where users encrypt their speech biometric templates using their own keys prior to enrolling them in the authentication server. During the matching stage, the the user again encrypts the freshly generated speech feature using it's own encryption key and send only the encrypted feature for matching. Since the server has got only the encrypted features, it has to perform the matching process in encrypted domain. Hence, this paper redesign the back-end matching process to support the encrypted domain processing.

This can be achieved via fully homomorphic encryption (FHE) techniques. The FHE was invented by Craig Gentry in 2009 \cite{Gentry} to perform both multiplication and addition in encrypted domain without the need to decrypt the data. The state-of-the-art FHE schemes are efficient and used to redesign several machine algorithms to process encrypted data. Therefore, if the speech features are encrypted with FHE, the server should be able to perform the verification without the need to decrypt the feature vectors. Hence, this paper proposes a methodology to exploit the properties of FHE scheme to develop a privacy-preserving speaker verification system in the encrypted domain.

\subsection{Notations}
We use bold lower-case letters like $\mathbf{x}$ to denote column vectors; for row vectors we use the transpose $\mathbf{x}^T$. We use bold upper-case letters like $\mathbf{A}$ to denote matrices, and identify a matrix with its ordered set of column vectors. Real numbers are denoted as $\mathbb{R}$and a real number matrix with $d \times d$  is denoted as $\mathbb{R}^{d \times d}$. We use $\mathds{Z}_q$ to denote the ring of integers  modulo $q$, $\mathds{Z}_q^{n\times m}$ to denote the set of $n\times m$ matrix with entries in $\mathds{Z}_q$. An integer polynomial ring with degree $N$ is denoted as $\mathbb{Z}_q[X]/(X^N+1)$ where the coefficients of the polynomial is always bounded by $\mathbb{Z}_q$. $||\mathbf{x}||$ denotes the vector norm where    $||\mathbf{x}||=\sqrt{\mathbf{x}^T\mathbf{x}}$.

\subsection{Paper organisation}
The rest of this paper is organised as follows: state-of-the-art works related to the proposed scheme are summarised in Section \ref{Section: RelatedWork}. Building blocks required for the proposed work is provided in Section \ref{Section: Background}. Section \ref{S: Proposed Scheme} proposes the privacy-preserving speaker verification system using CKKS homomorphic encryption. Testing environment, data-set, and parameter selection to achieve $128-$bit security are provided in Section \ref{Section: Performance Analysis}. Experimental results and efficiency compared with the traditional scheme are given in Section \ref{Section: Experiments}. The security and privacy analysis is given in Section \ref{Section: Security and Privacy Analysis}  followed by conclusions are discussed in Section \ref{Section: Conclusion}.

\section{Related Works}\label{Section: RelatedWork}
Several services are now exploiting unique features of speech for healthcare monitoring \cite{Parkinson,PTSD,Alzheimer}, authenticating banking applications \cite{banking}, and smart home applications\cite{smarthome}. These services  need to collect and store users' speech data over the Internet. At the same time, privacy regulations like GDPR in Europe are enforcing organisations to provide sufficient privacy guarantee when they use, process and store customer data. Since speech data is considered as unique and contain personnel information, the privacy of the voice data should be guaranteed.

To achieve this, we require novel techniques to redesign the speech processing back-end systems to protect the privacy while ensuring the utility of the data. There are several privacy-preserving techniques in literature that transform various types of data into  encrypted domain using traditional homomorphic encryption or randomisation techniques i.e., facial biometric \cite{Erkin_2009,Rahul_TDSC_2016}, emotions \cite{Rahul_TDSC,Rahul_TAFF,Rahul_Bio}, or voice biometric \cite{ PathakRaj2013}.   

In the domain of speech processing, there are only a few notable privacy-preserving  works exist  \cite{Smaragdis2007, PathakRaj2013,  PorteloRaj2014, Rahul_TASL}. Smaragdis and Shashanka proposed the first application of secure multi-party computation (SMC) concepts for privacy-constrained speech technology \cite{Smaragdis2007}. In their work, they realised secure speech recognition using the hidden Markov model (HMM) and a generalised version of the Paillier public-key scheme, which allowed training and classification between multiple parties and achieved perfect accuracy. 

Pathak et. al redesigned  Gaussian Mixture Model (GMM) based speaker recognition \cite{PathakRaj2013} to achieve a similar privacy goal. This work relies on homomorphic cryptosystems such as BGN and Paillier encryption. This work has shown a proof-of-concept of privacy-preserving speaker recognition without compromising the accuracy. However, the shortcoming of the above cryptographic approaches is that far too much time is spent on the encryption i.e., few minutes required for processing.
 
Recently, the work in \cite{Rahul_TASL} used randomisation technique from information theory to develop a privacy-preserving speaker verification scheme. This work is neither computationally inefficient nor compromises privacy. The solution presented in \cite{Rahul_TASL} is significantly advanced than the existing solution in terms of accuracy, privacy and speed. However,  \cite{Rahul_TASL} is interactive and requires multiple rounds of computations and it cannot be used for different front-end systems. Moreover, the security of all the schemes mentioned above relies on a mathematically intractable problems such as integer factorisation (schemes based on randomisation) and discrete logarithm (schemes based on homomorphic encryption). As we started to see the rise of quantum computers, the security of all these might be broken soon \cite{RegevLWE2005}.

In contrast to the traditional partial homomorphic encryption schemes (i.e., Paillier, BGN, etc), the rise of fully homomorphic schemes (FHEs) show promising results recently in terms of efficiency. While FHE resists attacks arising from quantum computers (due to lattice hard problems \cite{RegevLWE2005}), they also support non-interactive computation on encrypted domain.  Some of the notables works in the intersection of FHE and machine learning are \cite{IBM, CryptoNet,DiNN} and many more. The work in \cite{IBM}, trains 30,000 logistics regression models  in encrypted domain within 20 minutes but performs encrypted domain inference in 5 seconds using CKKS FHE scheme. The work in \cite{CryptoNet}, jointly done by Princeton University and Microsoft in 2016, transforms a trained Convolutional Neural Network (CNN) into a model suitable for encrypted domain inferencing. The work uses a simple CNN with 5 layers and 28x28 input dimension for MNIST dataset and requires 400MB bandwidth and 5 minutes to perform inference in encrypted domain. Finally the work in \cite{DiNN}, uses a novel discretization approach to transform neural networks suitable for advanced FHE scheme. A simple Neural network with 3 layers (with hidden layer of 100 neurons) took only 1.7 seconds to perform image classification at 96\% accuracy for 128-bit security \cite{DiNN}. There are several other works in this domain that are focusing on redesigning the traditional machine learning (mainly deep learning) algorithms to work on FHE domain.

However, to the best of our knowledge, there are no FHE based speech processing machine learning algorithms exist in literature to achieve end-to-end privacy in real-time. Within this context, we develop a novel algorithm that changes the back-end of speaker verification system to process an encrypted speech data in real-time without the need for multiple rounds of communications. Moreover, the proposed algorithm supports real-time end-to-end encrypted speaker verification  for negligible loss of accuracy at 128-bit security.

\section{Background Information} \label{Section: Background}
This section briefly describes the four building blocks required for the proposed algorithm.

\subsection{The Speaker Verification Systems}\label{Section: SpeakerVerificationSystems}
As shown in Fig. \ref{Fig: BlockDiagram}, the speaker verification systems composed of two components: 1) front-end and 2) back-end. The front-end is mainly focused on extracting feature vectors from speech. The back-end performs noise reduction and similarity calculation of speaker features.

The front-end extracts a  number of acoustic features such as  linear predictive cepstral coefficients,  perceptual linear prediction coefficient, and  mel-frequency cepstral co-efficients. Then several techniques used to enhance these features to get a better verification accuracy. In 1995, Reynolds et al. \cite{Reynolds1995} applied Gaussian Mixture Model technique based on Universal Background Model (GMM-UBM) on these features to increase the accuracy by a significant percentage. Since then the GMM-UBM based speaker verification became the foundation of speaker verification research. Fifteen years later, Dehak et al. \cite{Dehak} proposed a ground breaking model called i-Vector to further decrease the speaker and channel variation while increasing the verification accuracy. Moreover, the i-Vectors are significantly lower dimension (i-Vectors are around 200x1 size) than the GMM-UBM models (GMM-UBM super-vectors are around 40,000x1 size). 

Recently, motivated by the powerful feature extraction capability of deep neural networks (DNNs), a lot of deep learning based speaker recognition methods were proposed \cite{DNNUBM,DNNGMM}. The DNN based schemes boost the performance of the speaker verification to a new level even in the wild environment. Similar to the i-Vector, the DNN based feature extraction methods output x-Vector \cite{xVector} and d-Vector \cite{dVector}. The dimensions of these vectors are very similar to  i-Vectors.

As depicted in Fig. \ref{Fig: BlockDiagram}, the front-end can either use GMM-UBM or DNN-UBM to obtain i-, x- or d-Vectors \cite{DNNUBM,DNNGMM}. Hence, only these features are sent to the server for enrolment and matching. This paper focuses on protecting these feature vectors stored and processed in the back-end. One of the dominant techniques used in the back-end to perform similarity calculation is Cosine distance between the enrolled (or claimed) and test feature vectors of the user \cite{Dehak,xVector,dVector}. 

\subsection{Cosine Distance Calculation}
Lets suppose, the user enrolled a  feature vector $\mathbf{w}_{\mathrm{target}}\in \mathbb{R}^{d \times 1}$ at the server. During the verification, the user is sending $\mathbf{w}_{\mathrm{test}}\in \mathbb{R}^{d \times 1}$. Now the server calculate the cosine distance between the target and test vectors as follows: 
\begin{equation}\label{Eqn: CDS}
\mathrm{score}(\mathbf{w}_{\mathrm{target}}, \mathbf{w}_{\mathrm{test}}) = \frac{<\mathbf{w}_{\mathrm{target}}, \mathbf{w}_{\mathrm{test}}>}{||\mathbf{w}_{\mathrm{target}}|| ||\mathbf{w}_{\mathrm{test}}||} \gtreqless \theta,
\end{equation}
where dimension $d$ is the size of the i-, d- or x-Vectors (the $d$ is between $150$ and $200$ in the state-of-the-art works). To further reduce the channel-and speaker depended noise, a  projection matrix $\mathbf{P}$ is used as follows \cite{Dehak}:
\begin{eqnarray}
\nonumber &&\mathrm{score}(\mathbf{w}_{\mathrm{target}}, \mathbf{w}_{\mathrm{test}})\\ 
\nonumber &=& \frac{(\mathbf{P}^T\mathbf{w}_{\mathrm{target}})^T(\mathbf{P}^T\mathbf{w}_{\mathrm{test}})}{\sqrt[]{(\mathbf{P}^T\mathbf{w}_{\mathrm{target}})^T(\mathbf{P}^T\mathbf{w}_{\mathrm{target}})}\sqrt[]{(\mathbf{P}^T\mathbf{w}_{\mathrm{test}})^T(\mathbf{P}^T\mathbf{w}_{\mathrm{test}})}},\\
\label{Eqn: CDS2} &=& \frac{\mathbf{w}_{\mathrm{target}}^T\mathbf{Q}\mathbf{w}_{\mathrm{test}}}{\sqrt[]{\mathbf{w}_{\mathrm{target}}^T\mathbf{Q}\mathbf{w}_{\mathrm{target}} . \mathbf{w}_{\mathrm{test}}^T\mathbf{Q}\mathbf{w}_{\mathrm{test}}}} \gtreqless \theta,
\end{eqnarray}
where $\mathbf{Q}=\mathbf{P}\mathbf{P}^T \in \mathbb{R}^{d\times d}$.

\subsection{Fully Homomorphic Encryption}
Fully Homomorphic Encryption (FHE) schemes support homomorphic properties such as addition and multiplication in encrypted domain. To explain this briefly, lets denote two numbers in plain domain as $a$ and $b$ and the corresponding homomorpically encrypted values in encrypted domain as $[a]$ and $[b]$. Denote the encryption and decryption functions as $Enc(.)$ and $Dec(.)$. The encryption function takes $a$ and $b$ in plain domain and public key $pk$ as inputs and outputs the corresponding encrypted value i.e., $Enc(a,pk) = [a]$ and $Enc(b,pk) = [b]$. The decryption function takes the encrypted value and secret key $sk$ as inputs and outputs the plain domain values i.e., $Dec([a],sk) = a$ and $Dec([b],sk) = b$.
Within this context, FHE properties allows to compute addition and multiplication in encrypted domain without the need to decrypt the value i.e., $[a] + [b] = Enc(a+b,pk)$, $[a].[b] = Enc(a.b,pk)$. Therefore, mathematical functions can be computed in encrypted domain using only encrypted values. For example, if a cloud wants to compute a function $f(x,y) = x^2+2xy+3y^3$ but only has encrypted inputs $[x]$ and $[y]$, the cloud can exploit the FHE to evaluate the function as follows: $f([x],[y]) = [x].[x] + 2.[x].[y] + 3.[y].[y].[y]$ where $Dec(f([x],[y]),sk) = f(x,y)$. Since the cloud is not holding the secret key $sk$, the evaluated function $f([x],[y])$ remains in encrypted domain.

An encryption scheme with the above FHE properties was invented by Craig Gentry in 2009 \cite{Gentry}. The scheme is based on Lattice-based cryptography hence secure against attacks arising from quantum computers \cite{Ajtai1996,RegevLWE2009,RahulIoT}. Since Gentry's ground breaking work, there are numerous improvements were made by several researchers to improve efficiency and scalability. Currently FHE has reached an inflection point where several relatively complex algorithms can be evaluated in encrypted domain in near-real time \cite{CryptoNet,DiNN,IBM}. Single-instruction-multiple-data (SIMD) is one of the powerful techniques that has enhanced the efficiency of FHE by more than 3 orders of magnitude \cite{SIMD}. While there are handful of FHE schemes, this paper focuses on FHE scheme based  Cheon-Kim-Kim-Song (CKKS) \cite{CKKS} since it is the most efficient method to perform approximate homomorphic computations over real and complex numbers.

\subsection{CKKS FHE Scheme}
CKKS scheme supports real numbers and SIMD operation, therefore, its a suitable candidate for applications rely on vectors of real-numbers. CKKS works with polynomials because they provide a good trade-off between security and efficiency as compared to standard computations on vectors. 

Given a message $\mathbf{m} \in \mathbb{R}^M$, a vector of real values, it is first encoded into a plaintext integer polynomial $m(\mathbf{X}) \in \mathbb{Z}[X]/(X^N+1)$ where $M=N/2$ and $N$ denotes the degree of the polynomial. The CKKS encryption encrypts $m(\mathbf{X})$ into two ciphertext polynomials $(c_0(\mathbf{X}) \in \mathbb{Z}_q[X]/(X^N+1) , c_1(\mathbf{X}) \in \mathbb{Z}_q[X]/(X^N+1)$ where $q$ is the size of the ciphertext modulo. In ciphertext domain, CKKS supports homomorphic addition, multiplication, and rotation operations. The rotation
operation homomorphically performs a cyclic shift of the vector by some step. The multiplication and
rotation operations in the CKKS scheme need additional corresponding
evaluation keys and the key-switching procedures.

Moreover, each real number data is scaled with some big integer $2^\Delta$, called the scaling factor, and then rounded to an integer prior to encrypting the data. When the two data encrypted with the CKKS scheme are multiplied homomorphically, the scaling factors of the two data are also multiplied. This scaling factor should be reduced to the original value using the rescaling operation (i.e., dividing by $2^\Delta$).

In CKKS, the size of the ciphertext are big (i.e., $q$ is big) hence it requires higher computational complexity. To reduce the complexity, the residue number system can be used. In the residue number system, the big integer is split into several small integers, and the addition and the multiplication of the original big integers are equivalent to the corresponding component-wise operations of the small integers i.e., $2^q = 2^{q_0 + q_1 + q_2 + \ldots + q_L+q_{L+1}}$ where $q_0 > q_i, i=1,\ldots L$, $q_{L+1} > q_i, i=1,\ldots L$, and $\Delta \approx q_1 \approx q_2 \approx q_3 \approx \ldots q_L$. The $L$ denotes the number of multiplications can be performed to a  ciphertext correctly. For example, if there are four CKKS ciphertexts $[a_1]$, $[a_2]$, $[a_3]$, and $[a_4]$ then  $[a_1].[a_2]$ requires one level of multiplication and $[a_1].[a_2].[a_3]$ requires two levels of multiplications. Instead of performing  $[a_1].[a_2].[a_3].[a_4]$ via three multiplications,  computing  $x= [a_1].[a_2]$ and  $y=[a_3].[a_4]$ followed by $x.y$ will require only two levels of multiplications. The efficiency of an algorithm is depend on circuits with smaller multiplicative depths.

The security of the CKKS scheme relies on the polynomial degree $N$ and the ciphertext modulo $q$. Table \ref{Table: CKKS N vs q} shows the parameters for achieving 128-, 192- and 256-bit security. For a given $N$, the maximum size for $q$ is decreasing with the increasing security level. If the application requires more levels of multiplication in ciphertext domain then it requires larger $q$. For a given security model, only way to increase the size of $q$ is by increasing the size of $N$. The increasing the $N$ has consequences in terms of computational complexity. 

\begin{table}[]
\centering
\caption{CKKS parameters for different security levels.}
\label{Table: CKKS N vs q}
\begin{tabular}{|c|c|c|c|}
\hline
      & 128-bit Security & 192-bit Security & 256-bit Security \\ \hline
N     & Max. size of q   & Max. size of q   & Max. size of q   \\ \hline
1024  & 27               & 19               & 14               \\ \hline
2048  & 54               & 37               & 29               \\ \hline
4098  & 109              & 75               & 58               \\ \hline
8192  & 218              & 152              & 118              \\ \hline
16384 & 438              & 305              & 237              \\ \hline
32768 & 881               & 611              & 476              \\ \hline
\end{tabular}
\end{table}

\subsection{Newton-Rapshon Method for Inverse Square Root Calculation}\label{Section: Newton}
While FHE computes multiplication and addition in encrypted domain, several fundamental mathematical operations such as finding an inverse or a square root of a number is not feasible. However, we can use Newton iterative method introduced by Isaac Newton in 1669 \cite{Newton} to calculate these in a FHE friendly way. Since the cosine distance calculation in (\ref{Eqn: CDS2}) requires inverse square root operation, this section describes the Newton iterative method to perform this operation using just multiplication and addition.

Let's define a function $f(x) = \frac{1}{x^2} - a$, where the root of this function gives the inverse square-root of $a$ i.e., $\frac{1}{x^2} - a = 0$ leads to $x=\frac{1}{\sqrt{a}}$. The Newton iterative formula for finding the root is given by the following equation \cite{Initx_0}:

\begin{equation}\label{Eqn: Newton}
x_{n+1} = x_n - \frac{f(x_n)}{f'(x_n)},
\end{equation}
where $f'(x_n) = -\frac{2}{x_n^3}$ is the derivation of $f(x)$ at $x=x_n$. Hence, using this derivation, the equation (\ref{Eqn: Newton}) can be modified into:
\begin{equation}\label{Eqn: Newton2}
x_{n+1} = 1.5x_n - 0.5ax_n^3.
\end{equation}
To find an inverse square root of $a$, the equation (\ref{Eqn: Newton2}) must be repetitively computed. The number of iteration required is heavily dependent on $x_0$ i.e., the initial value for (\ref{Eqn: Newton2}). If the $a$ is bounded by $a_{min}$ and $a_{max}$ (i.e., $a_{min} < a < a_{max}$) then a good starting point is the average of the bounds i.e., $x_0 = 0.5 (\frac{1}{\sqrt{a_{min}}} +  \frac{1}{\sqrt{a_{max}}})$ \cite{Initx_0}. With this initialisation, (\ref{Eqn: Newton2})  can be computed using only multiplications, hence, it is FHE friendly replacement for inverse square root operation.

 \section{The Proposed Scheme}\label{S: Proposed Scheme}
 In this section, we put together all the techniques explained in Section \ref{Section: Background} to develop a privacy-preserving speaker verification technique using CKKS based fully homomorphic encryption scheme. The user will be provided with a client application to  extract features from their speech, generate secret and public keys required for CKKS FHE scheme, and interact with the server.

 \subsection{Feature Extraction}
 As shown in Fig. \ref{Fig: BlockDiagram}, the speech data can be converted into a feature vector with dimension $d$. Regardless of the feature extraction models, the dimension of the feature vector is around $200$. The raw speech data goes through several speech processing modules to get Mel frequency cepstral coefficients (MFCC) followed by GMM supervectors with large dimension. These high dimensional vectors can be reduced via several advanced techniques such as i-Vector models (GMM/UBM i-Vectors), d- and x- vectors via Deep Neural Networks (GMM/UBM DNN) . Since this work focuses on the back-end of the speaker verification system, we selected a computationally efficient GMM/UBM based i-Vector model for feature extraction. The proposed scheme is directly applied to any front-end feature extraction scheme that outputs a low-dimensional vector (i.e., $d$ is around $200$). 
 
 \subsection{Key generation for CKKS FHE scheme}
 The security key generation relies on several factors and depends on the underlying application. As shown in Table \ref{Table: CKKS N vs q}, the high-level parameters $N$ and $q$ must be selected by considering the efficiency and security. Moreover, scaling factor $2^\Delta$ and number of multiplication levels $L$ must be set in advance.  Since the application might be used by several users, the server presets these parameters common for all the users. Given these global parameters ($N$, $q$, $\Delta$ and $L$), each user (i.e., the client application running on the user's device) generates public-key and secret-key. The public key can be used for encryption, rescaling and rotation and will be sent to the server. The secret key never leaves the user device.
 
 \subsection{Enrolling the feature vector}
Using the client application, the user can extract speech features, generate keys for encryption, and start the enrolment process. The enrolment process is simple and require executing the following 4 steps:

\begin{itemize}
  \item Extract a speech feature vector $\mathbf{w}_1 \in \mathbb{R}^{d\times 1}$ from speech
  \item Obtain the initialisation variable $x_0$ (more details about this will be provided in the next section)
  \item Generate and store secret-public key pairs ($sk$ and $pk$)
  \item Apply CKKS encryption to get the following encrypted vectors $[1.5x_0\mathbf{w}_1]$ and $[0.5x_0^3\mathbf{w}_1]$
\end{itemize}

Now user sends  $\{ID_u, pk, [1.5x_0\mathbf{w}_1], [0.5x_0^3\mathbf{w}_1] \}$  to the server for enrolment where $ID_u$ denotes the user ID. The server stores the data in a database against the user ID  $ID_u$.

 \subsection{Speaker verification}\label{Section: Speaker Verification ED}
 The speaker verification part is the core contribution of this proposed work. Similar to the enrolment, the user extracts a feature vector $\mathbf{w}_2 \in \mathbb{R}^{d\times 1}$ and applies CKKS encryption to get $[\mathbf{w}_2]$. For the encryption, the user uses the same key that is being generated during the enrolment stage. To complete the verification stage, the user sends $\{ID_u, [\mathbf{w}_2]\}$ to the server. Now server retrieves the stored data from database using $ID_u$ and  evaluate (\ref{Eqn: CDS2}) to obtain the verification score. The projection matrix $\mathbf{Q}$ in  (\ref{Eqn: CDS2}) is available to the server in plain domain. Please note that the matrix $\mathbf{Q}$ is obtained by the server during the training process and it doesn't derived from the user's speech data (see \cite{Dehak,Rahul_TASL} for more details).
 
 If we closely look at verification equation (\ref{Eqn: CDS2}), the server computes the numerator to get a scalar, then computes the denominator to get a scalar followed by division between these scalars. Hence, we can reformulate (\ref{Eqn: CDS2}) as follows:
 \begin{equation}\label{Eqn: CDS3}
\mathrm{score}([\mathbf{w}_1],[\mathbf{w}_2]) = \frac{1}{\sqrt{[a]}}. [\mathbf{w}_1]^T\mathbf{Q}[\mathbf{w}_2]
 \end{equation}
where 
\begin{equation}\label{Eqn: denominator}
[a] = [\mathbf{w}_1]^T\mathbf{Q}[\mathbf{w}_1].[\mathbf{w}_2]^T\mathbf{Q}[\mathbf{w}_2].
\end{equation}
Since $[a]$ in (\ref{Eqn: denominator}) is encrypted, it's not possible to directly compute the inverse square-root of  $[a]$ required for (\ref{Eqn: CDS3}). Hence, we exploit the Newton-Rapshon method as explained in Section \ref{Section: Newton}. Newton-Rapshon method is iterative,  hence, using (\ref{Eqn: Newton2}), the approximated result after first iteration is given by:
\begin{equation}\label{Eqn: 1st approx}
    \frac{1}{\sqrt{[a]}} \approx_1  1.5x_0 - 0.5[a]x_0^3,
\end{equation}
and after second iteration:
 \begin{equation}\label{Eqn: 2nd approx}
    \frac{1}{\sqrt{[a]}} \approx_2  1.5(1.5x_0 - 0.5[a]x_0^3) - 0.5[a](1.5x_0 - 0.5[a]x_0^3)^3,
\end{equation}
and so on and so forth ($\approx_1$ and $\approx_2$ denote approximation after first and second iterations, respectively). Using the first approximated value in (\ref{Eqn: 1st approx}), we can get approximated value for (\ref{Eqn: CDS3}) as follows:
 \begin{equation}\label{Eqn: CDS4 Approx}
\mathrm{score}([\mathbf{w}_1],[\mathbf{w}_2]) \approx_1 (1.5x_0 - 0.5[a]x_0^3) [\mathbf{w}_1]^T\mathbf{Q}[\mathbf{w}_2].
 \end{equation}

\begin{table*}[t]\large
\centering
\begin{eqnarray}
 \label{Eqn: CDS4 Approx Expansion} \mathrm{score}([\mathbf{w}_1],[\mathbf{w}_2]) &\approx_1& 1.5x_0[\mathbf{w}_1]^T\mathbf{Q}[\mathbf{w}_2] - 0.5x_0^3[\mathbf{w}_1]^T\mathbf{Q}[\mathbf{w}_1].[\mathbf{w}_2]^T\mathbf{Q}[\mathbf{w}_2][\mathbf{w}_1]^T\mathbf{Q}[\mathbf{w}_2],
 \\
  \label{Eqn: CDS4 Approx Moving inside}
 &=&[ 1.5x_0\mathbf{w}_1]^T\mathbf{Q}[\mathbf{w}_2] - [0.5x_0^3\mathbf{w}_1]^T\mathbf{Q}[\mathbf{w}_1].[\mathbf{w}_2]^T\mathbf{Q}[\mathbf{w}_2][\mathbf{w}_1]^T\mathbf{Q}[\mathbf{w}_2],
 \\
 \label{Eqn: CDS4 Approx Levels}
&=&\underbrace{\underbrace{[ 1.5x_0\mathbf{w}_1]^T\mathbf{Q}}_{Level 1}[\mathbf{w}_2]}_{Level 2} - \underbrace{\underbrace{\underbrace{\underbrace{[0.5x_0^3\mathbf{w}_1]^T\mathbf{Q}}_{Level 1}[\mathbf{w}_1]}_{Level2}.\underbrace{\underbrace{[\mathbf{w}_2]^T\mathbf{Q}}_{Level1}[\mathbf{w}_2]}_{Level2}}_{Level 3}\underbrace{\underbrace{[\mathbf{w}_1]^T\mathbf{Q}}_{Level 1}[\mathbf{w}_2]}_{Level 2}}_{Level 4}.
 \end{eqnarray}
\end{table*}

Using (\ref{Eqn: CDS3}), we can expand (\ref{Eqn: CDS4 Approx}) into  (\ref{Eqn: CDS4 Approx Expansion}) (shown in the top of the next page). As described in Section \ref{Section: Newton}, $x_0$ in (\ref{Eqn: CDS4 Approx Expansion}) is the initialisation variable and its already supplied by the user to the server during the enrolment. Hence, equation  (\ref{Eqn: CDS4 Approx Expansion}) can be revised as (\ref{Eqn: CDS4 Approx Moving inside}). As shown in (\ref{Eqn: CDS4 Approx Levels}), the server requires four multiplication levels to compute  (\ref{Eqn: CDS4 Approx Moving inside}). Similarity, we can incorporate the second iterative result in (\ref{Eqn: 2nd approx}) which consumes six multiplicative levels. The result of third iteration consumes seven multiplication levels and nine levels for fourth iteration and so on and so forth. Increasing the number of multiplication levels lead to larger parameters for CKKS encryption which will directly impact the efficiency of the scheme. Given this context, lets focus on how server can compute the score using only the first iteration result as shown in (\ref{Eqn: CDS4 Approx Levels}).

The server first computes all four Level 1 multiplications shown in (\ref{Eqn: CDS4 Approx Levels}) which is a encrypted-vector-plain-matrix computation involving plain matrix $\mathbf{Q} \in \mathbb{R}^{d \times d} $.  The SIMD feature in CKKS supports element-wise multiplication and addition operation. To exploit this feature, the server needs to reassemble the matrix $\mathbf{Q} $ into $d$ vectors $\mathbf{q}_1$,  $\mathbf{q}_2$, $\ldots$,  $\mathbf{q}_d$ as described below. If
 \[
\mathbf{Q}
=
\begin{bmatrix}
    q_{11} & q_{12} & q_{13} & \dots  & q_{1d} \\
    q_{21} & q_{22} & q_{23} & \dots  & q_{2d} \\
    \vdots & \vdots & \vdots & \ddots & \vdots \\
    q_{d1} & q_{d2} & q_{d3} & \dots  & q_{dd}
\end{bmatrix}
\] then
 
\begin{eqnarray}
\nonumber \mathbf{q}_1 &=& [q_{11}, q_{22}, q_{33}, \ldots, q_{dd}],\\
\nonumber \mathbf{q}_2 &=& [q_{12}, q_{23}, q_{34}, \ldots, q_{d1}],\\
\nonumber \mathbf{q}_3 &=& [q_{13}, q_{24}, q_{35}, \ldots, q_{d2}],\\
\nonumber \vdots &=& ~~~~~~~~~~ \vdots,\\
\nonumber \mathbf{q}_d &=& [q_{1d}, q_{21}, q_{32}, \ldots, q_{dd-1}],
\end{eqnarray}
where diagonals of $\mathbf{Q}$ are reassembled as vectors in a cyclic manner in $\mathbf{q}_1,\ldots, \mathbf{q}_d$. Hence we can perform the vector-matrix computation $[\mathbf{w}]^T\mathbf{Q}$ as follows:
\begin{equation}\label{Eqn: vector-matrix mult}
[\mathbf{w}]^T\mathbf{Q}=[\mathbf{w}]\odot \sum_{i=1}^{d}\mathbf{q}_i,
\end{equation}
where $\odot$ denotes the element-wise Hadamard product operation. Since the vector $[\mathbf{w}]$ is encrypted, the result of (\ref{Eqn: vector-matrix mult}) is a $d$-dimensional encrypted vector. Moreover, in (\ref{Eqn: vector-matrix mult}), multiplications are element-wise hence the whole operation consumes only one CKKS multiplication level. 
 
All four Level 1 computations provide encrypted vectors which will be used for Level 2 computation. At Level 2, there are four multiplications which are encrypted-vector-encrypted-vector dot product computation. To perform this dot product computation, we exploit CKKS SIMD element-wise multiplication and rotation features. Lets suppose, $[\mathbf{a}] = [[a_1], [a_2], [a_3], [a_4], \ldots, [a_n]]$ and $[\mathbf{b}] = [[b_1], [b_2], [b_3], [b_4], \ldots, [b_n]]$, then to compute $[\mathbf{a}]^T[\mathbf{b}]$, we first perform element-wise Hadamard product using CKKS SIMD operation as follows: $[\mathbf{a}] \odot [\mathbf{b}] = [[a_1.b_1], [a_2.b_2], [a_3.b_3], [a_4.b_4], \ldots, [a_n.b_n]]$. To obtain the final answer, we repetitively shift the vector elements and perform addition. For example, if $n=4$ then we shift the vector by 2 elements and add as follows: 
\begin{eqnarray}
\nonumber &&[[a_1.b_1], [a_2.b_2], [a_3.b_3], [a_4.b_4]] \\
\nonumber &+&[[a_3.b_3], [a_4.b_4],[a_1.b_1], [a_2.b_2]]\\
\nonumber &=&[[a_1.b_1+a_3.b_3], [a_2.b_2+a_4.b_4], \ldots, [a_4.b_4+a_2.b_2]]
\end{eqnarray}
 Then we rotate the added vector by 1 element and add it again as follows:
 \begin{eqnarray}
\nonumber &&[[a_1.b_1+a_3.b_3], [a_2.b_2+a_4.b_4], \ldots, [a_4.b_4+a_2.b_2]]\\
\nonumber &+&[[a_2.b_2+a_4.b_4], [a_3.b_3+a_1.b_1], \ldots,[a_1.b_1+a_3.b_3]]\\
\nonumber&=& [[a_1.b_1+a_1.b_1+a_3.b_3+a_4.b_4],\ldots]
\end{eqnarray}
 Now the first element of the vector contains the correct answer for the dot product computation. One of the condition for this repeated rotation and addition is that the $n$ should be a power of two. This condition can be easily met by concatenating zeros at the end of the vectors.  We need to perform $log_2(n)$ repeated rotation and addition. Finally the dot product computation consumes only one multiplication level for the element-wise multiplication. The rotation and addition doesn't consume any multiplication level. 

The Level 2 computation described above produces an encrypted CKKS scalar (not vector). Now  for the Level 3 and Level 4, we only need to perform encrypted-scalar-encrypted-scalar multiplication which is straightforward to compute. Using these computations, we can obtain the approximated score in encrypted domain.

Now this encrypted score will be sent to the client application. The client application decrypts it and check if the score is above the threshold to authenticate the user. While this approach can be used for different applications (i.e., if the underlying application is about measuring the medical condition, then this score will represent the severity), this paper will only consider speaker verification use case.

\section{Parameter Selection and Performance Analysis}\label{Section: Performance Analysis}
This section describes the dataset used for the experiments, results and the complexity, security, and privacy analysis of the proposed algorithm.

\subsection{Parameter Selection}\label{Section: Parameter Selection}

We start with selecting parameters for the CKKS encryption. In this experiment, we stick with $128-$bit security. We select three sets of parameters as shown in  Table \ref{Table: Parameters}. Set I considers the smallest possible $N$ suitable for the application. Since $q$ is limited to $218$ when $N=8192$, the maximum number of multiplication levels we can do is limited to 4 without loosing a lot of accuracy. Therefore, we will only use one Newton-Raphson iteration to find the inverse square root. If we set the base prime size and the special prime size to 41 i.e., $q_1=q_{5}=41$, then we are left with $218 - 82= 136$-bit. We split this into four 34-bit required for the four multiplication levels. Since we are using all the available bits, the security of Set I is 128-bit.

\begin{table}[h]
\centering
\caption{Three sets of security parameters.}
\label{Table: Parameters}
\begin{tabular}{|c|c|c|c|}
\hline
                                                                                                              & \begin{tabular}[c]{@{}c@{}}Set I\\ N=8192\end{tabular} & \begin{tabular}[c]{@{}c@{}}Set II\\ N=16384\end{tabular} & \begin{tabular}[c]{@{}c@{}}Set III\\ N=16384\end{tabular} \\ \hline
No. Iterations                                                                                                & 1                                                      & 1                                                        & 2                                                         \\ \hline
No. Multiplication Levels                                                                                     & 4       & 4                     & 6                     \\ \hline
\begin{tabular}[c]{@{}c@{}}Maximum size for $q$\\ from Table \ref{Table: CKKS N vs q}\end{tabular} & 218     & 438                   & 438                   \\ \hline
$q_0$                                                                                                         & 41      & 60                    & 60                    \\ \hline
$\Delta$                                                                                                      & 34      & 40                    & 40                    \\ \hline
Used size of $q$                                                                                              & 218     & 280                   & 360                   \\ \hline
Security                                                                                                      & 128-bit & \textgreater{}128-bit & \textgreater{}128-bit \\ \hline
\end{tabular}
\end{table}

Set II and Set III use higher order polynomial with degree $N=16384$. This supports maximum size of $q=438$, which gives a lot of flexibility on prime sizes and multiplication levels. Set II considers only one Newton-Rapshon iteration, hence four multiplication levels are required. We set high bits sizes for base prime, special prime and scales $\Delta$ (i.e., 60, 60, and 40, respectively), totalling only 280 bits which is smaller than the allowable 438 bits. Therefore the security of Set II is higher than 128-bit. 

To increase the accuracy of finding inverse square root of encrypted number, we need to go for the result of second Newton-Rapshon iteration which requires 6 multiplication levels. The parameters for this is shown in Set III in Table \ref{Table: Parameters}. Similar to Set II, the security of Set III is higher than 128-bit.

\subsection{The dataset}
TIMIT speech corpus  has been used to evaluate the accuracy and reliability of the proposed algorithm \cite{TIMIT}. The TIMIT speech corpus contains broadband recordings (each recording lasts for around 3 seconds) of $630$ speakers of eight major dialects of American English. Each speaker has $10$ speech samples. Out of $10$ samples, $8$ were used to extract feature vector for enrolment. We use GMM/UBM based i-Vector for the experiments. However, as described earlier, the proposed model can be used for DNN/UBM based x- or d- vector speaker verification systems. 

For experiment, we follow the same approach used in \cite{Rahul_TASL} as a baseline. In \cite{Rahul_TASL}, the TIMIT data corpus has been split into two: 1) the first two dialect regions with $151$ speakers are used for testing and 2) the last four dialect region with $277$ speakers were used to build background model. Table \ref{Table: TIMIT} shows the statistics of the TIMIT dataset.
\begin{table}[!h]
\centering
\caption{Statistics of TIMIT database.}
\label{Table: TIMIT}
\begin{tabular}{|c|c|c|c|}
\hline
\begin{tabular}[c]{@{}c@{}}Dialect\\ Region (DR)\end{tabular} & \#Male       & \#Female     & \textbf{Total} \\ \hline
DR1                                                        & 31           & 18           & \textbf{49}    \\ \hline
DR2                                                        & 71           & 31           & \textbf{102}   \\ \hline \hline
DR3                                                        & 79           & 23           & \textbf{102}   \\ \hline
DR4                                                        & 69           & 31           & \textbf{100}   \\ \hline \hline
DR5                                                        & 62           & 36           & \textbf{98}    \\ \hline
DR6                                                        & 30           & 16           & \textbf{46}    \\ \hline
DR7                                                        & 74           & 26           & \textbf{100}   \\ \hline
DR8                                                        & 22           & 11           & \textbf{33}    \\ \hline
\textbf{Total}                                           & \textbf{438} & \textbf{192} & \textbf{630}   \\ \hline
\end{tabular}
\end{table}
Since, $8$ speech samples from the $151$ speakers are used for enrolling the user in server, the remaining $2$ samples per user have been used for verification. Initially we perform the following two baseline tests in plain domain using (\ref{Eqn: CDS2}):
\subsubsection{1. Genuine Attempts:- Client-Client} In this test, for each speaker, the score is calculated using the speaker's enrolled data against the speaker's two test utterances. Hence, the scores for $151\times 2 =302$ tests are obtained using  (\ref{Eqn: CDS2}).
\subsubsection{2. Imposter Attempts:- Imposter-Client} In this test, each speaker's test utterances are tested against other $150$ users' entolled feature vector. This leads to $151\times (151-1)\times 2=45300$ tests  and the score for each test has been obtained  using (\ref{Eqn: CDS2}).  

Before we present the results, let us define False Acceptance Rate (FAR), False Rejection Rate (FRR) and Accuracy.
\begin{itemize}
\item FAR = $\frac{\textrm{No. False Acceptance}}{\textrm{Total No. Imposter Attempts}} \times 100\%$,
\item FRR = $\frac{\textrm{No. False Rejection}}{\textrm{Total No. Geinune Attempts}}\times 100\%$,
\item Accuracy= $\frac{\textrm{No. Correct Acceptance and Correct Rejection}}{\textrm{Total No.  Attempts}} \times 100\%$,
\end{itemize}
where FAR and FRR are the two types of errors and False Acceptance means the system grants access to an impostor, and False Rejection means the system denies access to an enrolled speaker. From FRR and FAR, we can get Equal Error Rate (EER). EER represents the operating point where the FAR is equal to the FRR. 
\begin{figure}\centering
 \frame{ \includegraphics[trim={0cm 0 0cm 0},clip,width=3.5in]{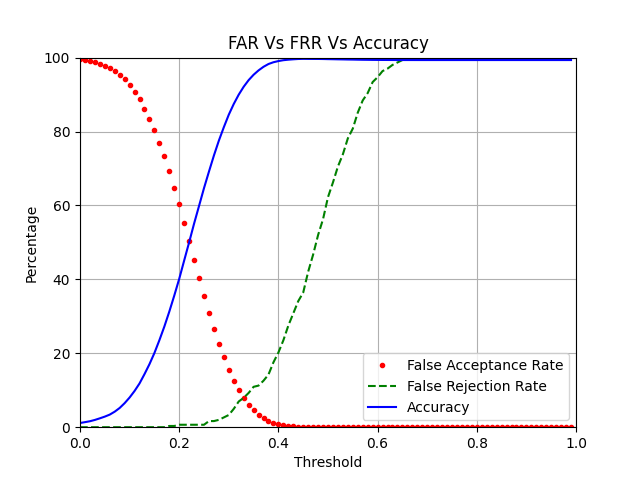}}
\caption{Baseline results showing False Acceptance Rate (FAR), False Rejection Rate (FRR), and Accuracy in plain domain.}
\label{Fig: Plain Accuracy}
\end{figure}
Using these definition, we can present the baseline results as shown in Fig. \ref{Fig: Plain Accuracy}. Since number of imposter attempts are significantly higher than the genuine attempts, the Accuracy curve in Fig. \ref{Fig: Plain Accuracy} might be misleading (i.e., it's approaching $100\%$ as it rejects large number of imposter attempts). Hence, we will stick to EER to compare the performance. The EER of the baseline model is around $7.8\%$ when the threshold is $0.34$. In the following section, we analyse the proposed scheme.

\section{Experimental Results}\label{Section: Experiments}
We implement the proposed algorithm in Python using TenSEAL library \cite{Tenseal} to interact with the C++ SEAL FHE library\footnote{https://github.com/Microsoft/SEAL}. The source code of our implementation can be found here: https://github.com/rahulay1/iVectorTenSEAL/tree/master. We essentially repeat the same steps that we used to evaluate the baseline model. We tested all 3 sets of CKKS parameters shown in Table \ref{Table: Parameters}. We compare the time requirements using a high end and medium end laptops. For the high end, we use a Razor laptop with 16GB ram and 6 cores (12 CPUs) with upto 4.1GHz speed. This can be treated as server. For the medium end, we use a MacBook Pro laptop with 8GB ram and 2 cores (4 CPUs) with upto 2.5GHZ speed. The specification of the medium end laptop is comparable to the specifications of medium end smartphones (i.e., Samsung Galaxy A Series phones), hence, can be considered for running the client application.

\subsection{Initialisation of Newton-Rapshon parameter}
\begin{figure}\centering
  \frame{\includegraphics[trim={0cm 0 0cm 0},clip,width=3.5in]{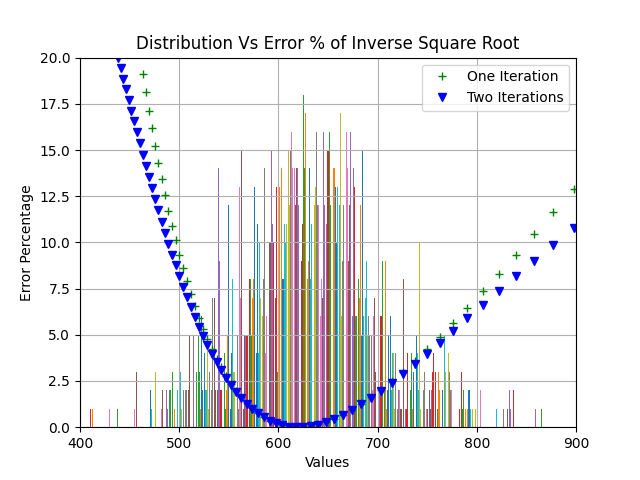}}
\caption{Distribution of $a$ in (\ref{Eqn: denominator}) using 0.7 million data points. The convex curves denote the relative error percentage of calculating the inverse square-root of the underlying data in the distribution using Newton-Rapshon method compared to the actual value.}
\label{Fig: Newton}
\end{figure}

Before we start the experiment, its very important to initialise the variable $x_0$ in (\ref{Eqn: Newton2}) for Newton-Rapshon method. As explained in Section \ref{Section: Newton}, if $x_0$ is closer to the actual inverse-square root, then the convergence is much faster. Therefore, finding the distribution of $a$ within this context is important. According to (\ref{Eqn: denominator}), $a = \mathbf{w}_1^T\mathbf{Q}\mathbf{w}_1.\mathbf{w}_2^T\mathbf{Q}\mathbf{w}_2$ and we can obtain the distribution of this value using the TIMIT dataset. Using all 630 speakers we could obtain more than 0.7 million sample values for $a$ i.e.,($629\times630\times2$). Using these samples, we plot the distribution of $\frac{1}{\sqrt{a}}$ in Fig. \ref{Fig: Newton}. From this, we can clearly see that $x_0$ should be initialised between 400 and 900. Instead of initialising the average of 400 and 900, we initialised as $0.45(400+900)=650$ as bigger chunk of data is around $650$.

Now we calculate the inverse square-root of $a$ using the iterative approach and compare it with the actual answer in the same Figure Fig.    \ref{Fig: Newton}. The two convex graphs shows the relative error percentage of iterative approach compared to actual value (i.e., $\frac{|iterative- value - actual-value|}{actual-value}\times 100$) for the first and second iterations. While there are no significant differences between the first and second iterations, the error is less than $2.5\%$ for bigger chunk of $a$. Therefore, we can safely use the Newton-Rapshon method to compute the inverse square-root of encrypted number as we proposed. We will evaluate the loss of EER due to this approximation in the next section.

\subsection{EER loss comparison of the proposed scheme against baseline approach}
\begin{figure}\centering
  \frame{\includegraphics[trim={0cm 0 0cm 0},clip,width=3.5in]{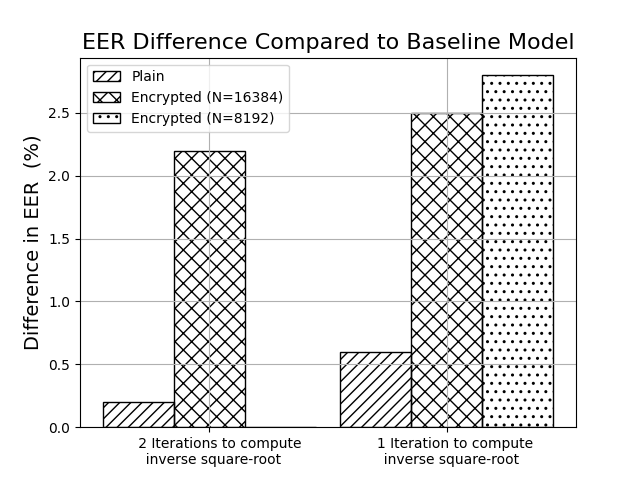}}
\caption{The difference in EER of the proposed method against the baseline model.}
\label{Fig: EER Comparision}
\end{figure}

Before we start experimenting the encrypted speaker verification algorithm proposed in Section \ref{Section: Speaker Verification ED}, we need to check the loss of EER when we replace the actual inverse square-root function with Newton-Raphshon iterative function. The result of this experiment is presented in Fig. \ref{Fig: EER Comparision} (see the first and third bars in Fig. \ref{Fig: EER Comparision}).  Use the result of 2 iterations of Newton-Rapshon method leads to $0.2\%$ loss in EER while 1 iteration leads to loss of $0.6\%$ EER compared to the baseline EER. Thanks to the careful selection of the initialisation value, these EER losses are negligible. 

Now we can compare the results of the proposed encrypted speaker verification system. These results are depicted in the remaining three bars of Fig. \ref{Fig: EER Comparision} (second, fourth and fifth bars). These bars correspond to the CKKS parameters in Set-III, Set-II, and Set-I  in Table \ref{Table: Parameters}, respectively. With Set-III parameters ($N=16384$ with 2 iterations), the loss of EER compared to baseline approach is around $2.2\%$. For other two sets, the EER losses are $2.5\%$ and $2.8\%$, respectively. The main reason for this is due to approximation  scaling factor $\Delta$ of the CKKS scheme. Since Set I uses small  $\Delta$ compared to Set-II, the experiment based on Set-I loses more precision of the underlying values hence lose in EER compared to Set-II. Nevertheless, $2.8\%$ loss in EER is not significant when we consider the time required for this verification is near-real time as discussed below. 

\subsection{Computational time and processing requirements}
One of the challenges that hinders the adoption of FHE in real application is it's ability to perform computation  in real-time. As we discussed in the literature review section, FHE has reached an inflection point where real-time application can be implement fully using FHE schemes. The result of the proposed scheme also support this statement. For all three sets of CKKS parameters in Table \ref{Table: Parameters}, we measured the time required to complete the key generation, enrolment, verification and decryption. These results are presented in Table \ref{Table: Time}. Table \ref{Table: Time} depicts results for four sets of experiments for each CKKS parameter, totalling 12 experiments.

For each CKKS parameter set, the experiment was conducted in high-end (C1) and medium-end (C2) laptops for two different feature dimensions ($d=100$ and $d=200$). From Table \ref{Table: Time}, we can observe that significant amount of time is spent on generating public and secret keys. However, this is one time effort and can be done in offline. The other three operations impact the real-time performance. The time required to encrypt a speech template (noted as Enrol)  is between 11ms and 55ms. As expected, the time consuming operation is verification.  For Set-I CKKS parameters, the verification can be done within 1.3 seconds (the EER loss of this set is $2.8\%$). For Set-III, while EER loss is limited to $2.2\%$, the time required to perform verification on C1 laptop is around 7 seconds which may be suitable for near-real-time application. The most efficient operation is decryption and require between 1ms and 12 ms for all 12 experiments.

If we consider a typical scenario where users uses a medium-end hardware and the server uses high-end hardware then the total delay due to FHE scheme could be 11ms + 1.2 seconds + 2ms  $<$ 1.3 seconds which is suitable for many real-time applications such as mobile banking, healthcare monitoring, etc.

When it comes to processing power, key generation, encryption and decryption do not require much CPU power (refer to the screenshot in Fig. \ref{Fig: CPU RAM usage}). However, almost $100\%$ of the available processing power will be consumed by the verification. Since the verification involves several vector dot products, these can be highly parallelized to exploit all the CPUs. As shown in Fig. \ref{Fig: CPU RAM usage}, all 12 CPUs in C1 is being used to complete the verification. Since C2, has only four CPUs, its performance is almost 3 times slower than C1 (refer Table \ref{Table: Time}).

\begin{table*}[]
\centering
\caption{Absolute time (seconds) required for the different steps in the proposed algorithms. The rows denoted by C1 are the results of running the proposed algorithm on a A Razor Laptop with 16GB ram and 6 cores (12 CPUs) with upto 4.1GHz speed and the rows of C2 are the results of running the proposed algorithm on a MacBook Pro laptop with 8GB ram and 2 cores (4 CPUs) with upto 2.5GHZ speed. The algorithm is tested for two different feature dimensions i.e., $d=100$ and $d=200$. Notations: KG - Key Generation, Dec. - Decryption and Veri. - Verification.}
\label{Table: Time}
\begin{tabular}{|c|c|cccc|cccccccc|}
\hline
\multirow{4}{*}{}   & \multirow{4}{*}{Dim.} & \multicolumn{4}{c|}{\begin{tabular}[c]{@{}c@{}}N=8192\\ Inverse Square-root from Iteration 1\end{tabular}}     & \multicolumn{8}{c|}{N=16384}                                                                                                                                                                                         \\ \cline{3-14} 
                    &                       & \multicolumn{3}{c|}{\multirow{2}{*}{Client}}                                         & \multirow{2}{*}{Server} & \multicolumn{4}{c|}{Inverse Square-root from Iteration 1}                                                           & \multicolumn{4}{c|}{Inverse Square-root from Iteration 2}                                      \\ \cline{7-14} 
                    &                       & \multicolumn{3}{c|}{}                                                                &                         & \multicolumn{3}{c|}{Client}                                                           & \multicolumn{1}{c|}{Server} & \multicolumn{3}{c|}{Client}                                                           & Server \\ \cline{3-14} 
                    &                       & \multicolumn{1}{c|}{KG}    & \multicolumn{1}{c|}{Enrol} & \multicolumn{1}{c|}{Dec.}  & Veri.                   & \multicolumn{1}{c|}{KG}     & \multicolumn{1}{c|}{Enrol} & \multicolumn{1}{c|}{Dec.}  & \multicolumn{1}{c|}{Veri.}  & \multicolumn{1}{c|}{KG}     & \multicolumn{1}{c|}{Enrol} & \multicolumn{1}{c|}{Dec.}  & Veri.  \\ \hline
\multirow{2}{*}{C1} & d=200                 & \multicolumn{1}{c|}{2.282} & \multicolumn{1}{c|}{0.016} & \multicolumn{1}{c|}{0.001} & 1.247                   & \multicolumn{1}{c|}{12.775} & \multicolumn{1}{c|}{0.052} & \multicolumn{1}{c|}{0.007} & \multicolumn{1}{c|}{6.442}  & \multicolumn{1}{c|}{12.760} & \multicolumn{1}{c|}{0.052} & \multicolumn{1}{c|}{0.004} & 6.858  \\ \cline{2-14} 
                    & d=100                 & \multicolumn{1}{c|}{2.467} & \multicolumn{1}{c|}{0.017} & \multicolumn{1}{c|}{0.001} & 0.684                   & \multicolumn{1}{c|}{12.875} & \multicolumn{1}{c|}{0.052} & \multicolumn{1}{c|}{0.004} & \multicolumn{1}{c|}{3.504}  & \multicolumn{1}{c|}{12.704} & \multicolumn{1}{c|}{0.052} & \multicolumn{1}{c|}{0.001} & 3.557  \\ \hline
\multirow{2}{*}{C2} & d=200                 & \multicolumn{1}{c|}{5.861} & \multicolumn{1}{c|}{0.011} & \multicolumn{1}{c|}{0.002} & 4.305                   & \multicolumn{1}{c|}{35.708} & \multicolumn{1}{c|}{0.043} & \multicolumn{1}{c|}{0.012} & \multicolumn{1}{c|}{19.737} & \multicolumn{1}{c|}{40.972} & \multicolumn{1}{c|}{0.045} & \multicolumn{1}{c|}{0.009} & 19.875 \\ \cline{2-14} 
                    & d=100                 & \multicolumn{1}{c|}{5.629} & \multicolumn{1}{c|}{0.011} & \multicolumn{1}{c|}{0.002} & 2.156                   & \multicolumn{1}{c|}{36.060} & \multicolumn{1}{c|}{0.042} & \multicolumn{1}{c|}{0.009} & \multicolumn{1}{c|}{9.759}  & \multicolumn{1}{c|}{40.028} & \multicolumn{1}{c|}{0.041} & \multicolumn{1}{c|}{0.004} & 9.855  \\ \hline
\end{tabular}
\end{table*}
\subsection{Analysis of storage, memory and bandwidth requirements}
\begin{figure}\centering
  \frame{\includegraphics[trim={0cm 0 0cm 0},clip,width=3.5in]{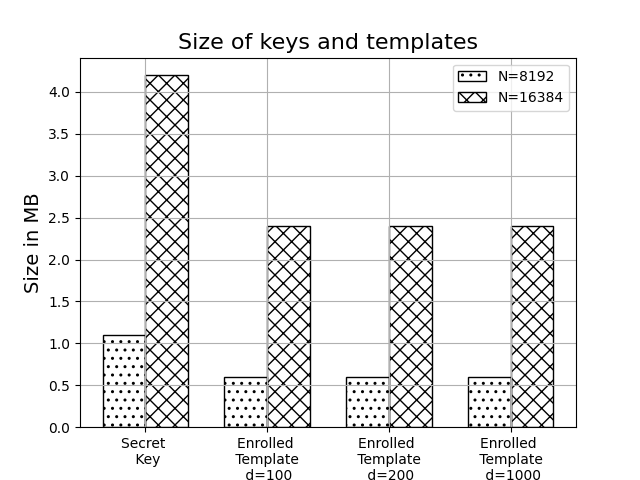}}
\caption{Comparing the storage requirements for secret keys and encrypted templates. Another element which is not provided in this figure is public-key (110MB and 756MB for N=8192 and N=16384, respectively).}
\label{Fig: Size Comparision}
\end{figure}

In the proposed scheme, both the client and server need storage, memory and communication bandwidth to exchange data between them. The client on user device must keep the secret key and share the pubic key with the server who keeps the public key for verification. Moreover, the client needs to share the encrypted templates with the server during the enrolment and verification. Hence, the server needs more storage to keep the encrypted templates. Fig. \ref{Fig: Size Comparision} shows the storage requirements for several of the components discussed above. The storage required for both secret and templates are less than 5MB for both the order of polynomial degree considered for CKKS scheme. However, higher order polynomial require higher storage. Since these polynomials can contain several slots for input vector, there is no difference on storage when the feature dimension $d$ increases from 100 to 1000. The dominant element that require a large storage is the public key (110MB for $N=8192$ and 0.75GB for $N=16384$).  Since this key must be communicated to the server, we also need relatively high bandwidth during the enrolment process. The main reason for this is that these public keys contains several keys for rescaling and rotation operations in the encrypted domain.

\begin{figure*}\centering
 \frame{ \includegraphics[trim={0cm  1.5cm 0cm 0cm},clip,width=7in]{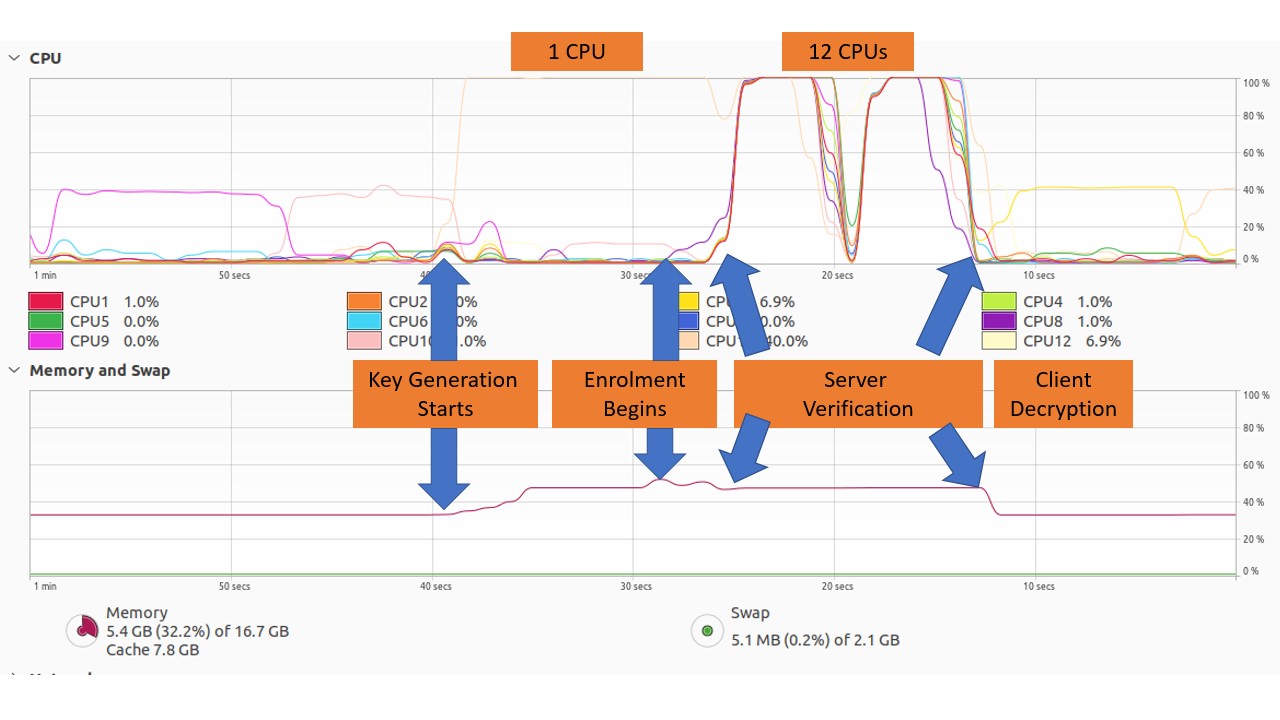}}
\caption{The screenshot is showing the usage of processing units and RAM in C1 laptop. Only about 20\% of the RAM (2.5GB) is consumed during the process, however, 100\% of all 12 CPUs are used during the verification. This phenomena clearly shows that the proposed algorithm is bounded by CPUs.}
\label{Fig: CPU RAM usage}
\end{figure*}

The usage of RAM is shown in Fig. \ref{Fig: CPU RAM usage}. When the process started, only about 20\% of the total available memory is being used in contrast to the usage of CPUs. Hence the efficiency of the proposed algorithms (mainly the use of FHE) is not dominated by the available memory. Hence, medium-end devices with up to 4GM RAM is sufficient to run the client.

\section{Privacy and Security Analysis}\label{Section: Security and Privacy Analysis}
This section analyses the privacy of the stored speech features followed by the security of the whole system.

\subsection{Privacy Analysis}
The aim of the proposed algorithm is to stop the server from learning the result of the inference. The proposed algorithm exploits CKKS FHE scheme where public key is being used for encryption, rescaling and rotation operations while secret key is being used for decryption. The proposed scheme requires only public keys to be sent to the server and secret key is never leave the user's device. Therefore, the server cannot obtain the inference results.

Another potential privacy vulnerabilities is the identity linkage attack i.e., if the users enrol their biometrics in multiple services, the service providers might collude and profile the users using the similarities of the speech features. However, this attack is not possible because the CKKS encryption is probabilistic hence, the server cannot distinguish the encrypted messages even if they contain the same message and encrypted using the same keys \cite{RegevLWE2009}. As long as the users' secret keys are protected, it is infeasible for the rogue service providers to profile the users. 

\subsection{Security Analysis}
While it is infeasible to decrypt the CKKS ciphertext without secret keys, there might be other ways the system can be compromised. For example, the attacker might have stolen the user device with secret keys or the attacker compromised the encrypted templates stored on the server or obtain the speech recording of the user.  In this section we investigate each of this scenario and show how the proposed scheme mitigates the security vulnerabilities.

\subsubsection{Compromised user device attacks}
In this attack, the adversary has access to the user device and the CKKS parameters
stored during the enrolment. But do not have access to the user's speech to generate legitimate speech feature. Hence, the adversary tries to combine the parameters from the compromised user device with the features of other users. Then the adversary
tries to verify against the compromised user’s enrolled template residing at the server. To evaluate this, 2 × 150 × 151 tests [300 test utterances from other users are combined with the parameters of the compromised user device and this is repeated for all the users] are conducted and the corresponding decision scores are obtained. Essentially the result of this experiment is already presented by the EER loss comparison in Fig. \ref{Fig: EER Comparision}. The loss of $2.8\%$ EER means that the FAR curve in  Fig. \ref{Fig: Plain Accuracy} is shifted by $2.8\%$ leading to accepting 28 more false claims for every 1000 impostor attempts. However, this can be reduced by using a small threshold for verification which will impact the FRR.

\begin{figure}\centering
  \frame{\includegraphics[trim={0cm 0 0cm 0},clip,width=3.5in]{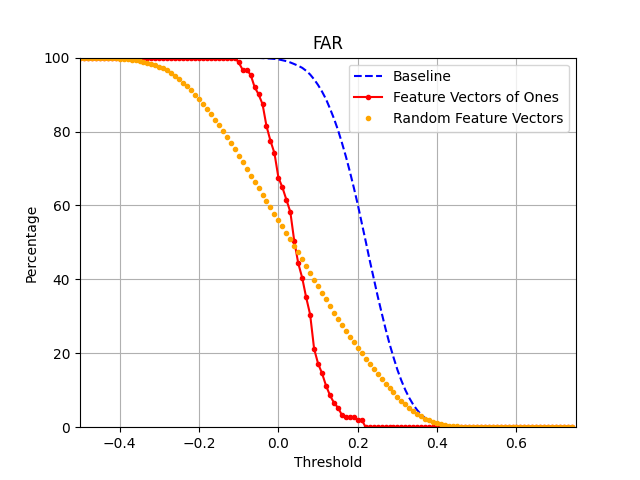}}
\caption{False acceptance rate reduces when an attacker uses random feature vectors or patterned feature vectors for verification.}
\label{Fig: Random Attacks}
\end{figure}

Sometimes the adversary generates completely a random feature vector or a patterned feature vector to maximise the score. For the patterned feature vector, we generated a vector with ones. Now these artificial feature vectors can be encrypted and decrypted by the stolen credential and used to conduct speaker verification. The result of this experiment is shown in Fig. \ref{Fig: Random Attacks}. The FAR of these attacks are lower than the baseline approach hence the adversary is worse off with these attacks.

\subsubsection{Compromised server attacks}
In this attack, the adversary has access to the enrolled encrypted data and public keys of all the users stored at the server. Hence, the adversary might attempt to modify the encrypted templates using FHE properties. Or he might use those encrypted templates during the verification process. However, none of these attacks will succeed without the secret keys. Moreover, if there is a compromise, the users can re-enrol using different set of  public and secret keys. Since compromised speech vectors are encrypted, they can be revoked (similar to passwords) even though the underlying data is biometric and unique to the user.

\subsubsection{Compromised user voice attacks}
In this attack, the attacker has access to the user’s voice recording but does not
have access to the parameters stored at the user device. Now the attacker generates random public and secret key pairs and tries to impersonate. The success of this attack is equivalent to breaking the CKKS FHE scheme hence this attack is also infeasible.

\section{Conclusion}\label{Section: Conclusion}
This paper presents a novel algorithm to process encrypted speech features using fully homomorphic encryption suitable for real-time speaker verification systems. The proposed algorithm exploits fully homomorphic encryption for arithmetic of approximate numbers (aka CKKS scheme) to achieve 128-bit   security against classical and quantum computers. To measure the performance, a well known speech corpus was used to conduct rigorous experiments. The end-to-end encrypted privacy-preserving scheme only requires  1.3 seconds to complete the verification in FHE domain. The accuracy in terms of equal-error-rate, the proposed scheme is off by only 2.8\%. Privacy analysis shows that the proposed scheme mitigates the privacy vulnerabilities such as tracking and profiling that exists in traditional system. Moreover, the proposed scheme is secure and the system cannot be exploited without accessing the secret keys.
\balance


\begin{thebibliography}{1}

\bibitem{Parkinson} J. Rusz et al., “Smartphone allows capture of speech abnormalities associated with high risk of developing Parkinson’s disease,” IEEE Trans.
Neural Syst. Rehabil. Eng., vol. 26, no. 8, pp. 1495–1507, Aug. 2018.

\bibitem{PTSD} R. Xu et al., “A voice-based automated system for PTSD screening and
monitoring,” in Proc. of Med. Meets Virtual R XII,  pp. 552–558, 2012.

\bibitem{Alzheimer} Y. Yamada, K. Shinkawa, and K. Shimmei, “Atypical repetition in daily
conversation on different days for detecting Alzheimer disease: Evaluation
of phone-call data from a regular monitoring service,” JMIR Ment. Health,
vol. 7, no. 1, Art. no. e16790, Jan., 2020.

\bibitem{Dementia} D. Shibata, S. Wakamiya, K. Ito, M. Miyabe, A. Kinoshita, and E.
Aramaki, “Vocabchecker: Measuring language abilities for detecting early
stage Dementia,” in Proc. Int. Conf. Intell. User Interfaces Companion,
 pp. 1–2, 2018.

\bibitem{JainAK}
A. K., Jain, and  K. Nandakumar, "Biometric Authentication: System Security and User Privacy," {\em Computer}, vol. 45, no. 11, 87-92, 2012.



\bibitem{Reynolds1995} 
D.A. Reynolds, R.C. Rose, "Robust text-independent speaker identification using Gaussian mixture speaker models,"{\em IEEE Trans.  Speech and Audio Processing} , vol.3, no.1, pp.72,83, Jan 1995.

\bibitem{DNNGMM}
Z. Bai, and X. L. Zhang, "Speaker recognition based on deep learning: An overview. Neural Networks2, 140, pp.65-99, 2021.

\bibitem{Dehak}
N. Dehak, P.J. Kenny, R. Dehak, P. Dumouchel, and P. Ouellet, "Front-end factor analysis for speaker verification", IEEE Trans. Audio, Speech, and Language Processing, 19(4), pp.788-798, 2011.

\bibitem{DNNUBM} 
Y. Lei, N. Scheffer, L. Ferrer, and M. McLaren, "A novel scheme for speaker recognition using a phonetically-aware deep neural network. In IEEE Int'l Conf. Acoustics, Speech and Signal Processing (ICASSP) (pp. 1695-1699), May, 2014.


\bibitem{xVector} 
D. Snyder, D. Garcia-Romero, D. Povey, and S. Khudanpur, "Deep neural network embeddings for text-independent speaker verification", In INTERSPEECH, pp. 999–1003, 2017.


\bibitem{dVector} 
E. Variani, X. Lei, E. McDermott, I.L. Moreno, and J. Gonzalez-Dominguez, "Deep neural networks for small footprint text-dependent speaker verification. In IEEE Int'l Conf. Acoustics, Speech and Signal processing, pp. 4052-4056, 2014.

\bibitem{Newton} 
I. Newton. Methodus Fluxionem et Serierum Infinit, 1966.

\bibitem{Gentry} 
C. Gentry,  "A fully homomorphic encryption scheme", Ph.D. thesis, Stanford Universityhttps://crypto.stanford.edu/craig, 2009. 

\bibitem{Ajtai1996} 
M. Ajtai,  "Generating hard instances of lattice problems", In Proc. of the twenty-eighth Annual ACM symposium on Theory of computing (pp. 99-108), July 1996.

\bibitem{RegevLWE2009} 
O. Regev, "On lattices, learning with errors, random linear codes, and cryptography", Journal of the ACM (JACM), 56(6), p.34, 2009.

\bibitem{RahulIoT} 
Y. Rahulamathavan, S. Dogan, X. Shi, R. Lu, M. Rajarajan,and A. Kondoz, "Scalar product lattice computation for efficient privacy-preserving systems", IEEE Internet of Things Journal, 8(3), pp.1417-1427, 2020.



\bibitem{CryptoNet} 
N. Dowlin et al. “Cryptonets: Applying neural networks to
encrypted data with high throughput and accuracy” In Int’l
Conf. Machine Learning, 2016.

\bibitem{DiNN} 
F. Bourse, M. Minelli, M. Minihold,  and P. Paillier, "Fast homomorphic evaluation of deep discretized neural networks. In Annual Int'l Cryptology Conf. (pp. 483-512). Springer, Cham, Aug., 2018.

\bibitem{IBM} 
F. Bergamaschi et al. "Homomorphic Training of 30,000 Logistic Regression Models." International Conference on
Applied Cryptography and Network Security. Springer, Cham, 2019.

\bibitem{SIMD} 
N. P. Smart, and F. Vercauteren, "Fully homomorphic SIMD operations", Designs, codes and cryptography, 71(1), pp.57-81, 2014

\bibitem{CKKS} 
J. H. Cheon, A. Kim, M. Kim, and Y. Song. Homomorphic encryption for arithmetic of approximate numbers. In ASIACRYPT’17, pages 409–437, 2017.

\bibitem{TIMIT} 
J. Garofolo, et al. "TIMIT Acoustic-Phonetic Continuous Speech Corpus LDC93S1," Web Download. Philadelphia: Linguistic Data Consortium, 1993.

\bibitem{Rahul_TASL}
Y. Rahulamathavan, K. R. Sutharsini, I. G. Ray, R. Lu, and M. Rajarajan,. Privacy-Preserving iVector-Based Speaker Verification. IEEE/ACM Transactions on Audio, Speech and Language Processing (TASLP), 27(3), pp.496-506, 2019.

\bibitem{Tenseal}
A. Benaissa, B. Retiat, B. Cebere, A.E. Belfedhal, "TenSEAL: A Library for Encrypted Tensor Operations Using Homomorphic Encryption", Int'l Conf. Learning Representations, Workshop on Distributed and Private Machine Learning, 2021.

\bibitem{Initx_0}
P. Kornerup,  and J.M. Muller," Choosing starting values for certain Newton–Raphson iterations. Theoretical computer science, 351(1), pp.101-110, 2006.


\bibitem{banking}
A. Phipps, K. Ouazzane, and V. Vassilev,  "Your password is music to my ears: cloud-based authentication using sound"., 11th Int'l Conf. Cloud Computing, 2021.

\bibitem{smarthome}
H. Isyanto, A. Arifin,  and M. Suryanegara, "Design and implementation of IoT-based smart home voice commands for disabled people using Google Assistant. In IEEE Int'l Conf. Smart Technology and Applications (ICoSTA) (pp. 1-6), Feb., 2020.

\bibitem{Smaragdis2007} 
P. Smaragdis, and M.V.S. Shashanka, "A Framework for Secure Speech Recognition," in IEEE Int'l Conf. Acoustics, Speech and Signal Processing, vol.4, no., pp.IV-969,IV-972, 15-20 April 2007.                              
\bibitem{Goldreich}
O.~Goldreich, ``Secure multiparty computation'',
(working draft), available: http://www.wisdom.wei
zmann.ac.il/ oded/pp.html. (Sep. 1998)


\bibitem{PathakRaj2013} 
M. Pathak, and B. Raj, "Privacy-Preserving Speaker Verification and Identification Using Gaussian Mixture Models," {\em IEEE Trans. Audio, Speech, and Language Processing} , vol.21, no.2, pp.397-406, Feb., 2013.

\bibitem{Erkin_2009}
Z.~Erkin, M.~Franz, J.~Guajardo, S.~Katzenbeisser, I.~Lagendijk, and T.~Toft,
  ``Privacy-preserving face recognition,'' in {\em Proc. 9th International
  Symposium on Privacy Enhancing Technologies}, PETS '09, pp.~235--253, 2009.


\bibitem{Rahul_TDSC_2016} 
Y. Rahulamathavan,  and M. Rajarajan, Efficient privacy-preserving facial expression classification. IEEE Transactions on Dependable and Secure Computing, 14(3), pp.326-338., 2015.



\bibitem{Rahul_TDSC}
Y.~Rahulamathavan, R.~Phan, S.~Veluru, K.~Cumanan, and M.~Rajarajan,
  ``Privacy-preserving multi-class support vector machine for outsourcing the
  data classification in cloud,'' {\em IEEE Trans. Dependable Secure
  Computing}, vol.~11, no.~5, pp.~467--479, Sept., 2014.

\bibitem{Rahul_Bio}
Y.~Rahulamathavan, S.~Veluru, R.~Phan, J.~Chambers, and M.~Rajarajan,
  ``Privacy-preserving clinical decision support system using gaussian kernel
  based classification,'' {\em IEEE Journal of Biomedical and Health
  Informatics}, vol.~18, no.~1, pp.~56--66, Jan., 2014.

  \bibitem{Rahul_TAFF}
Y. Rahulamathavan, R. Phan, J. Chambers, and D. Parish, ``Facial expression
  recognition in the encrypted domain based on local fisher discriminant
  analysis,'' {\em IEEE Trans. Affective Computing}, vol.~4, no.~1,
  pp.~83--92, Jan.-Mar., 2012.

 
\bibitem{PorteloRaj2014}
J. Portêlo, B. Raj, A. Abad and I. Trancoso, "Privacy-preserving speaker verification using secure binary embeddings," {\em Information and Communication Technology, Electronics and Microelectronics (MIPRO),  37th International Convention on}, Opatija, 2014, pp. 1268-1272, 2014.



\bibitem{RegevLWE2005}
O. Regev, "On Lattices, Learning with Errors, Random Linear Codes, and Cryptography," In Proc. 37th ACM Symp. on Theory of computing (STOC), pages 84--93, 2005.




\end{thebibliography}
\end{document}